# Synthesis of Ultra-thin Potassium Tungsten Bronze Single Crystals with Optically Contrasting Domains and Resistive Switching

Abdulsalam Aji Suleiman[1,2,3,4*], Amir Parsi[2], Hafiz Muhammad Shakir[5+], Hamid Reza Rasouli[2++], Doruk Pehlivanoğlu[2], Talip Serkan Kasırga[1,6*]

[1]National Nanotechnology Research Center – UNAM, Ankara 06800, Türkiye

[2]Institute of Materials Science and Nanotechnology, Bilkent University, Ankara 06800, Türkiye

[3]Department of Engineering Fundamental Sciences, Sivas University of Science and Technology, Sivas 58000, Türkiye

[4]Sivas Cumhuriyet University Nanophotonics Application and Research Center CÜNAM, Sivas 58140, Türkiye

[5]Department of Physics, Bilkent University UNAM, Ankara 06800, Türkiye

[6]Department of Physics, Middle East Technical University, Ankara 06800, Türkiye

[+]Current Address: College of Science and Engineering, Hamad Bin Khalifa University, Education City, Qatar Foundation 34110, Doha, Qatar

[++]Current Address: Institute of Physical Chemistry, Friedrich Schiller University Jena, 07743, Jena, Germany

*Corresponding author e-mail: suleiman@sivas.edu.tr, kasirga@unam.bilkent.edu.tr

**Abstract**

Potassium tungsten bronzes ($K_xWO_3$) are nonstoichiometric oxides in which alkali ions, i.e., $K^+$, occupy one-dimensional tunnels of the hexagonal $WO_6$ framework, enabling coupled ionic–electronic transport. While their bulk and nanostructured forms have been studied extensively, controlled synthesis of single-crystalline mesoscale samples suitable for device fabrication has remained limited. Here, we report a solid–liquid–solid (SLS) growth strategy that yields high-quality $K_xWO_3$ nanobelts with thicknesses down to ~36 nm and lateral sizes exceeding 100 μm. The crystals display sharp optical domains arising from local variations in potassium occupancy, as confirmed by spatially resolved Raman spectroscopy and electron diffraction. Under applied bias, these domains vanish irreversibly, consistent with lateral redistribution of $K^+$ ions along the tunnels. Two-terminal devices fabricated from individual nanobelts exhibit reproducible bipolar switching with resistance ratios of 10–30, characteristic short-term and long-term plasticity under pulsed excitation, and switching energies of ~25 nJ. These results establish $K_xWO_3$ as a model tunnel-structured oxide for studying electric-field-driven alkali-ion migration, while also highlighting its potential for stable, analog resistive switching and iontronic memory applications.

**Keywords:** Potassium tungsten bronze; solid–liquid–solid (SLS) growth; alkali-ion migration; analog resistive switching; iontronic

**Introduction**

Stoichiometric deviations in crystalline solids typically introduce vacancies or interstitials, which govern their structural, optical, and electronic properties[1-4]. In oxides with the general formula $ABY_3$, where A and B are cations and Y is an anion, the $BO_6$ octahedra can arrange in multiple configurations that accommodate different amounts of the A cation. This structural flexibility enables the formation of nonstoichiometric compounds with composition-dependent functionalities[5-7]. A well-known example is the tungsten bronzes ($A_xWO_3$), first investigated in the 19th century[8]. In these materials, A may be an alkali metal, a transition metal, or a lanthanide, and the hexagonal phase remains stable for $0.19 \leq x \leq 0.33$, as established by Magnéli and others[9]. The structure is composed of corner-sharing $WO_6$ octahedra that form three- and six-membered rings, generating one-dimensional tunnels along the c-axis. These tunnels accommodate the A-site cations and provide a natural pathway for ionic motion, while tungsten vacancies can further distort the lattice into cubic or tetragonal polymorphs.

Most prior work on tungsten bronzes has focused on bulk powders, nanoflakes, and nanowires synthesized by a range of techniques[10-14]. Although useful for probing average crystallographic and electronic behavior, such morphologies often conceal microscopic inhomogeneities. Indeed, electron diffraction on $K_xWO_3$ has revealed that even micrometer-sized "single crystals" can host multiple distinct patterns, reflecting local structural variations[15]. Comparable nanoscale inhomogeneities are common in transition-metal oxides (TMO) that support ion migration and phase transitions. To uncover the intrinsic ionic dynamics of $K_xWO_3$, single-crystalline samples with dimensions below the characteristic domain size are therefore essential.

Ion intercalation in TMO has recently emerged as a mechanism for resistive switching, where applied fields drive cation migration and trigger structural transformations. In our recent works, K- and Na-intercalated manganese oxides, for example, electric-field-driven $K^+$ or $Na^+$ motion induces reversible phase changes between layered birnessite and spinel forms, producing conductivity modulation[16, 17]. Such ion-driven mechanisms offer an alternative to vacancy- or filament-based switching, which often suffer from variability and poor reproducibility[18, 19]. This has motivated increasing interest in open-framework oxides that enable controlled ionic transport. Within this class, the tunnel-structured hexagonal phase of $K_xWO_3$ is especially attractive, as its one-dimensional channels confine $K^+$ transport, providing a possibly predictable and reversible migration pathway without large-scale structural degradation[20].

In this study, we synthesize high-quality single-crystalline $K_xWO_3$ nanobelts via a solid–liquid–solid (SLS) growth process on c-cut sapphire. The belts, as thin as ~36 nm and up to 120 μm long, display striking optical domain contrast arising from variations in K occupancy, as confirmed by Raman spectroscopy and electron diffraction. The K occupancy variation is induced by non-uniform strain distribution on the nanobelts, which emerges during the cool down after the synthesis, due to the thermal expansion coefficient mismatch between sapphire and $K_xWO_3$. When an in-plane bias of 3 to 5 V is applied, the contrast gradually vanishes, and reproducible bipolar switching emerges. The devices exhibit a resistance ratio of up to 30, gradual analog conductance modulation under pulsed bias, and switching energies of ~25 nJ. Unlike filamentary oxides, where switching is abrupt and localized, $K_xWO_3$ responds uniformly and reversibly within its one-dimensional tunnels, establishing it as a model platform for studying alkali-ion migration and a promising basis for robust iontronic memory.

**Results and Discussion**

The growth of $K_xWO_3$ crystals was achieved through a solid–liquid–solid (SLS) process, with conditions guided by our custom-built real-time optical chemical vapor deposition (RTO-CVD) chamber, described in detail in our earlier work[21]. A schematic of the chamber is shown in Figure S1a in the supporting information (SI), highlighting precursor placement on a sapphire substrate and real-time visualization through an optical window. In situ optical microscopy provided time-lapse images of nucleation and crystal evolution (Figure S1b), enabling direct correlation between growth kinetics and product morphology. This feedback proved essential in refining the synthesis recipe, allowing us to reproducibly obtain high-quality crystals with improved thickness control and uniformity.

The crystal structure of hexagonal $K_xWO_3$ is depicted in **Figure 1**, where corner-sharing $WO_6$ octahedra form rings of three and six units that generate one-dimensional tunnels along the c-axis. These tunnels host K ions whose occupancy determines the stoichiometry and stability of the bronze lattice. Unlike layered or perovskite oxides, where alkali ions diffuse between sheets or cages, the tunnel geometry provides a rigid, anisotropic pathway for ion transport while preserving lattice connectivity[5]. This structural feature is central to both the growth behavior and the mixed electro-ionic properties discussed later.

Upon optimizing the growth conditions in the RTO-CVD chamber, the synthesis was transferred to a split-furnace CVD system, where space-confined geometry was used to promote thinner and more uniform crystals. For this process, the mixed tungsten trioxide ($WO_3$) and potassium iodide (KI) were dispersed on a c-cut sapphire substrate before being capped with a second sapphire to enforce confinement. The chamber was evacuated to $10^{-2}$ mbar, flushed with argon to maintain an inert atmosphere, and heated to 700 °C. At this temperature, KI melts, wets the surface, and dissolves into the $WO_3$ precursor, initiating the formation of $K_xWO_3$, while $I_2$ escapes as vapor. This confined SLS route proved critical for achieving nanobelts with thicknesses down to ~36 nm and lateral dimensions exceeding 100 μm (Figure S2).

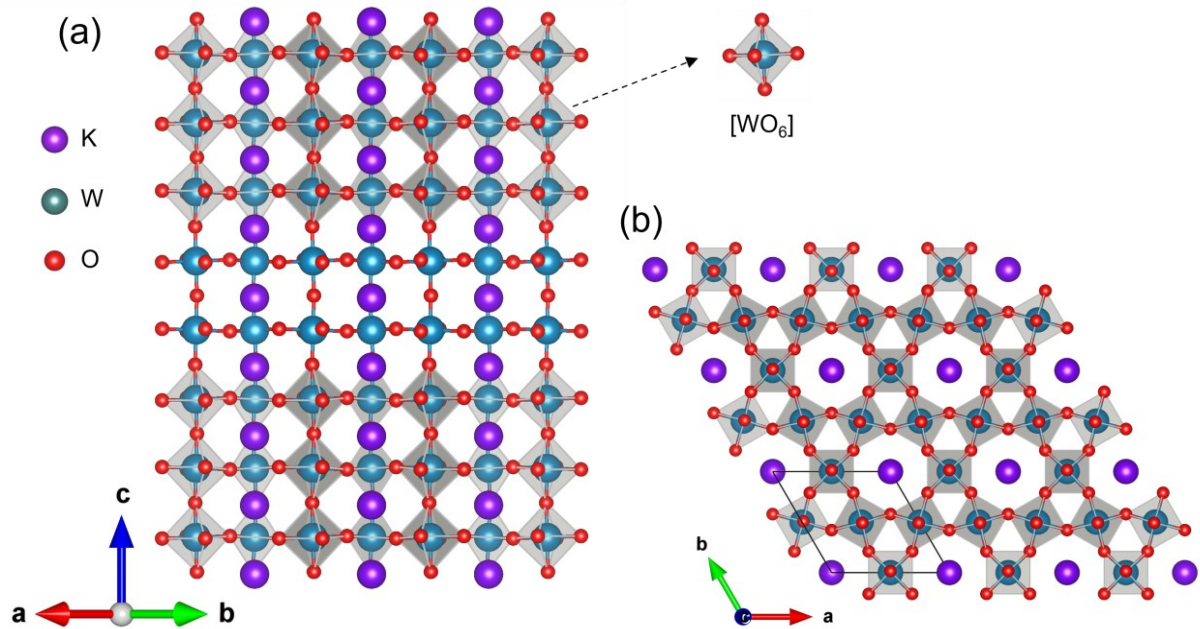

**Figure 1:** Crystal structure of hexagonal $K_xWO_3$. **(a)** Side view along the [100] direction showing the layered stacking of corner-sharing $WO_6$ octahedra, with $K^+$ ions occupying interlayer channels. **(b)** Top view along the c-axis, illustrating the hexagonal arrangement of $K^+$ ions within the $WO_3$ framework.

Upon inspection under OM, the as-grown $K_xWO_3$ single crystals displayed striking domain patterns, with alternating "bright" and "dark" regions clearly visible along their length **(Figure 2a)**. To probe these contrasts, we performed micro-Raman spectroscopy and mapping using a 532 nm excitation laser at a power of 100 μW. These domains—optically distinguishable as "dark" and "bright" regions—suggest inhomogeneous potassium distribution along the crystal. The Raman spectra from these domains **(Figure 2b)** display striking differences in both peak positions and linewidths, indicative of local structural and compositional variations. The Raman spectrum obtained from the dark domain closely resembles that of hexagonal $WO_3$, exhibiting strong vibrational modes associated with W–O stretching and bending[22]. Specifically, characteristic peaks around 259, 288, and 673 cm$^{-1}$, corresponding to bending (200–400 cm$^{-1}$) and stretching (650–850 cm$^{-1}$) modes of the $WO_6$ octahedra, align well with literature reports for the hexagonal tungsten bronze phase[23].

In contrast, the bright domain displays additional high-frequency peaks at ~917 and ~936 cm$^{-1}$, which are characteristic of terminal W=O stretching vibrations commonly observed in potassium-deficient tungsten bronzes[22]. Furthermore, the peak broadening and slight red/blue shifts observed in the Raman spectra—particularly in the dark domains—can be attributed to a combination of factors, including phonon confinement effects, local strain fields, and the presence of loosely bound potassium ions disrupting the periodic potential of the lattice. Variations in the oxidation state of W (e.g., presence of $W^{5+}$ alongside $W^{6+}$) may also contribute to the observed spectral asymmetry and band softening [24]. Raman intensity maps generated at representative wavenumbers—259, 795, 917, and 933 cm$^{-1}$ **(Figure**

**2c)**—clearly delineate the spatial distribution of vibrational modes and reveal abrupt interfaces between contrasting domains[25]. These maps strongly support the presence of homojunction-like boundaries arising from stoichiometric phase separation within the same crystal.

To further probe the structural stability and phonon dynamics across these domains, we conducted temperature-dependent Raman spectroscopy between 273 K and 410 K. The corresponding spectra and contour maps are presented in Figure S3. Across this temperature range, no emergence or disappearance of Raman modes was observed, indicating that the observed spectral differences are intrinsic and not thermally activated[26]. This also suggests that potassium redistribution is stable within this temperature window and that the phase contrast is not due to thermal fluctuations or phase transitions.

To investigate the vibrational anisotropy and lattice symmetry of $K_xWO_3$, we performed polarization-resolved micro-Raman spectroscopy. Polar plots of Raman intensity as a function of in-plane rotation angle for selected phonon modes are presented in Figure S4. These measurements allow us to probe the symmetry properties of individual vibrational modes by analyzing how their Raman intensities vary with light polarization. The angular response of the high-frequency peaks at 795, 917, and 936 cm$^{-1}$, associated with terminal W=O stretching modes, exhibits clear twofold symmetry, with maximum intensity at 90° and 270° under co-polarized configurations. This behavior is consistent with $A_1$-like modes vibrating along a principal crystallographic axis, likely the c-axis[27]. Their well-defined lobes and extinction angles support their assignment to symmetry-allowed Raman tensor components under backscattering geometry. The 259 cm$^{-1}$ mode, tentatively assigned to a bending or lattice mode involving K-ion displacements or W–O–W linkages, shows a weaker but still discernible anisotropic pattern, with lobes slightly offset from the high-frequency modes. This suggests a different symmetry origin or a mixed character involving off-axis displacements.

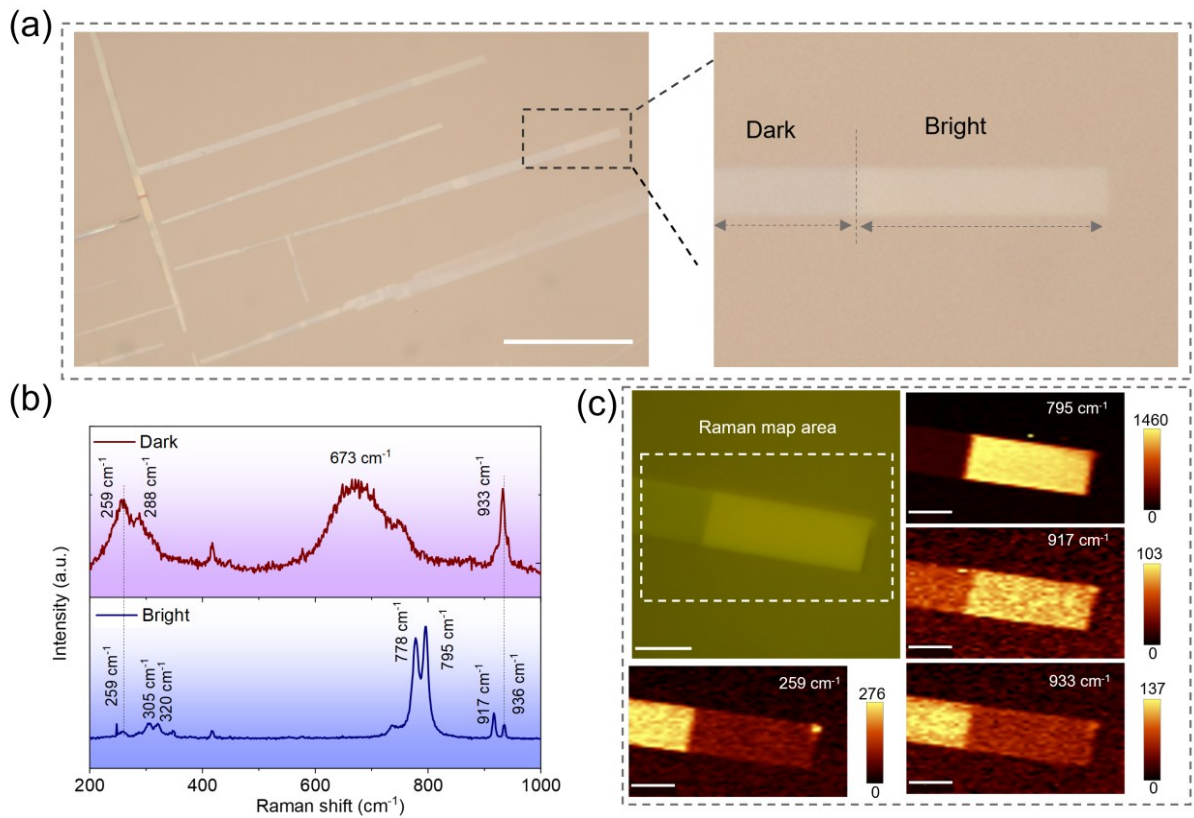

**Figure 2: (a)** Optical micrograph of $K_xWO_3$ flakes showing bright and dark domains. Scale bar: 20 μm. **(b)** Raman spectra from each domain, revealing distinct peak positions and widths. **(c)** Raman intensity maps at 259, 795, 917, and 933 cm$^{-1}$, highlighting phonon contrast across the domains. Scale bar: 5 μm.

Next, to elucidate the crystallographic structure of the synthesized $K_xWO_3$ crystals, we employed a combination of X-ray diffraction (XRD), high-resolution transmission electron microscopy (HRTEM), high-angle annular dark field scanning TEM (HAADF-STEM), and selected area electron diffraction (SAED). The XRD pattern in **Figure 3a** aligns well with the reference data (ICDD 01-070-1229), confirming the formation of a hexagonal tungsten bronze (HTB) phase. A prominent diffraction peak at low angle corresponds to the (110) plane, with an interplanar spacing of approximately 0.787 nm characteristic of the large tunnels in the HTB framework[5]. HRTEM image from the edge of the crystal shows a uniform crystalline structure of the sample in the observed field of view (**Figure 3c**).

The experimentally estimated crystallographic planes of (110) with 0.787 nm and (310) with 0.364 nm d-spacing labelled over the HRTEM image agree with our XRD results. These lattice spacings provide direct real-space confirmation of the HTB structure[28]. **Figure 3d** shows the results of the SAED pattern indexed along the [001] axis. This pattern is consistent with previously reported zone-axis diffraction for K-doped tungsten bronzes and supports a [001] projection of the HTB lattice[15]. The cross-sectional HAADF-STEM image shown in **Figure 3e** demonstrates the long-range ordering of the $WO_6$ octahedral framework, and the intensity profile along the crystal shows the spacing between W atoms. Together, these results confirm that the $K_xWO_3$ crystallizes predominantly in a hexagonal tungsten bronze framework with long-range order and local structural modulation, consistent with prior studies on alkali-doped tungsten oxides[15, 28-31].

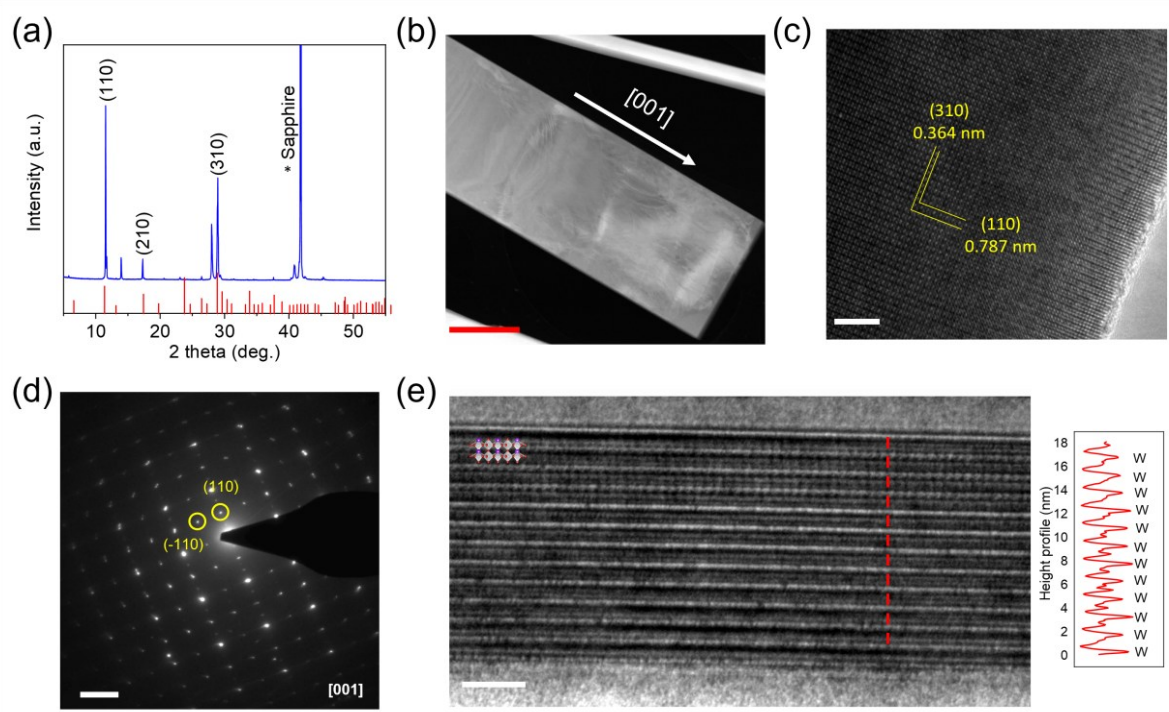

**Figure 3:** Structural characterization of hexagonal $K_xWO_3$ nanobelts. **(a)** XRD confirms phase purity and matches the hexagonal tungsten bronze structure (ICDD 01-070-1229). **(b)** TEM image of a representative single-crystalline belt oriented along the [001] direction. Scale bar: 500 nm. **(c)** HRTEM resolves lattice fringes corresponding to the (310) and (110) planes with spacings of 0.364 nm and 0.787 nm, respectively. Scale bar: 5 nm. **(d)** SAED pattern along the [001] zone axis showing sharp, well-defined spots consistent with hexagonal symmetry. Scale bar: 2 1/nm. **(e)** Cross-sectional HAADF-STEM reveals long-range ordering of the $WO_6$ octahedral framework, with the intensity profile confirming periodic lattice modulation across the belt thickness. Scale bar: 5 nm.

The elemental composition of the $K_xWO_3$ was first confirmed by energy-dispersive X-ray spectroscopy (EDX), as shown in **Figure 4a**, while the oxidation states of the constituent elements were examined by X-ray photoelectron spectroscopy (XPS). As shown in **Figure 4b**, the W 4f spectrum exhibits peaks at 37.3 eV ($4f_{5/2}$) and 34.9 eV ($4f_{7/2}$), characteristic of $W^{6+}$, while the shoulder on the $4f_{7/2}$ component indicates the presence of $W^{5+}$ state[10, 32]. Variations in the oxidation state of W suggest that some of the $W^{6+}$ cations are reduced to $W^{5+}$ to maintain charge balance in the structure, which is indicative of K incorporation during growth. **Figure 4c** displays the deconvolution for potassium, with the K $2p_{1/2}$ orbital at 294.9 eV and the K $2p_{3/2}$ orbital at 292.2 eV. Notably, no iodine-related peaks (e.g., I 3d) were detected as illustrated in Figure S5a, indicating that $I_2$ is fully removed during growth, and the K detected is attributed to K incorporation. The O 1s signal from the lattice is present but masked by the strong oxygen background from the sapphire substrate, as shown in Figure S5b.

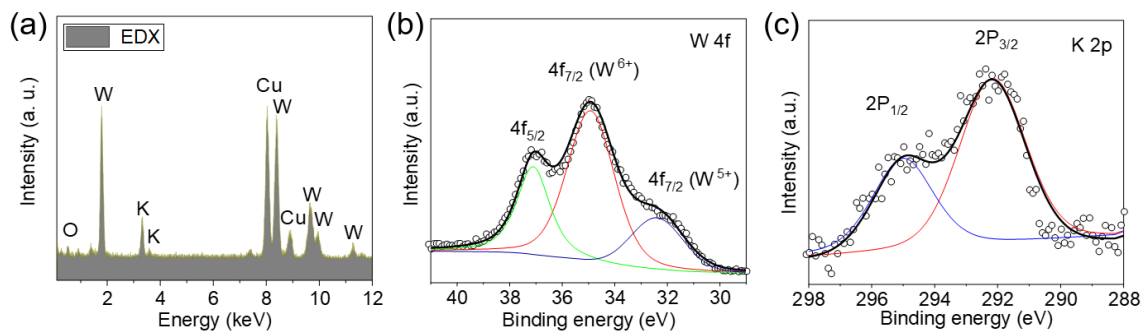

**Figure 4: (a)** EDX spectrum of $K_xWO_3$ nanobelts. The traces of Cu are due to the TEM grid. **(b)** XPS core-level regions of W 4f spectrum showing $4f_{7/2}$ and $4f_{5/2}$ peaks of $W^{6+}$ and $4f_{7/2}$ peak of $W^{5+}$. **(c)** XPS core-level regions of K 2p spectrum with $2p_{3/2}$ and $2p_{1/2}$ peaks confirming K incorporation.

The two-terminal electrical response of $K_xWO_3$ directly reflects its tunnel-confined alkali-ion dynamics. Two-terminal devices were contacted using molten indium probes, enabling us to probe pristine single crystals with minimal extrinsic effects[33]. **Figure 5a** shows the optical contrast in $K_xWO_3$ before and after electrical cycling. In the pristine state (top), the crystal exhibits pronounced bright stripes along the belt, while after repeated biasing (bottom), the contrast disappears and the crystal appears uniformly dark. This transformation reflects a field-driven redistribution of K ions within the tunnels, which homogenizes local stoichiometry. As this causes an increase in the overall crystal length, strain on the crystal increases. This strain is sometimes relieved by buckling of the crystal, as shown in the supporting information, consistent with lattice expansion along the c-axis during ion migration.

These observations indicate that the optical domains are not intrinsic defects but metastable patterns that collapse once the ionic distribution becomes more uniform under bias. The presence of such sharp domains can be explained by a mechanism induced by uniaxial strain on the crystals' c-axis. The crystals form at elevated temperatures, as discussed earlier. As the samples cool down to room temperature after the crystal nucleation, due to the difference between the thermal expansion coefficients of sapphire and $K_xWO_3$, a uniaxial strain is present on the crystal. Moreover, when the K concentration is high in a region of the crystal, its unit cell is longer along the crystal's c-axis. As a result, during the cool down, as the crystal is compressed, uniaxial strain leads to the expulsion of K to reduce the crystal length. This reduces the strain by creating K-rich and poor domains, at the energy cost of creating domain boundaries. Such domain formation is very similar to that observed in single-crystalline $VO_2$ nanobeams grown on $SiO_2$ substrates and strained ferroelectric systems[34-36]. Also, aging in ambient conditions further amplifies this strain relaxation via slow diffusion of K ions within the lattice. After two weeks, nanobelts exhibit pronounced buckling along the c-axis, as confirmed by optical and scanning electron microscopy (SEM) imaging (Figure S6).

As shown in **Figure 5b**, the I–V curves display stable bipolar hysteresis, accompanied by real-time optical contrast changes recorded during cycling (Figure S7). Unlike filamentary oxides such as $HfO_2$ or $TiO_2$, where switching is abrupt and stochastic, $K_xWO_3$ responds in a smooth and reproducible manner, reflecting mixed ionic–electronic conduction regulated by tunnel-confined $K^+$ motion[19, 37-39]. At a read voltage of 2 V, high-resistance states (HRS) on the order of $10^8$–$10^9$ Ω and low-resistance states (LRS) of ~$10^7$ Ω were maintained over repeated sweeps **(Figure 5c)**. The resistance switching ratio of up to 30, consistent with partial homogenization of the ionic distribution across the belt.

Most importantly, the devices support gradual, analog conductance updates that emulate synaptic functions. Paired pulses (4 V, 0.3 s) induce short-term depression, with the gradual conductance change from 2.94 to 2.88 nS **(Figure 5d)**. Long trains of positive and negative pulses produce long-term synaptic plasticity, with reproducible modulation between ~2.3 and 3.0 nS in discrete steps **(Figure 5e)**. Each update consumes ~25 nJ and occurs on timescales of 300 ms. These numbers are not competitive with filamentary memristors, but they are directly comparable to intercalation systems such as K–$MnO_2$ and Na–$MnO_2$, where alkali-ion migration and redox govern resistive states[40]. The difference is that in layered oxides, resistive switching often relies on costly structural transitions, while in $K_xWO_3$, the rigid tunnel framework confines the ions and preserves structural integrity. Compared with chalcogenides such as $Cu_2S$, which can deliver lower operating voltages but suffer from poor endurance or chemical instability, $K_xWO_3$ offers a more robust and reproducible oxide platform[41]. A broader benchmarking with other ion-electronic memristors is provided in Table 1 in the SI, emphasizing the distinct role of tunnel-confined ion transport in enabling gradual, reproducible, and non-filamentary resistive switching.

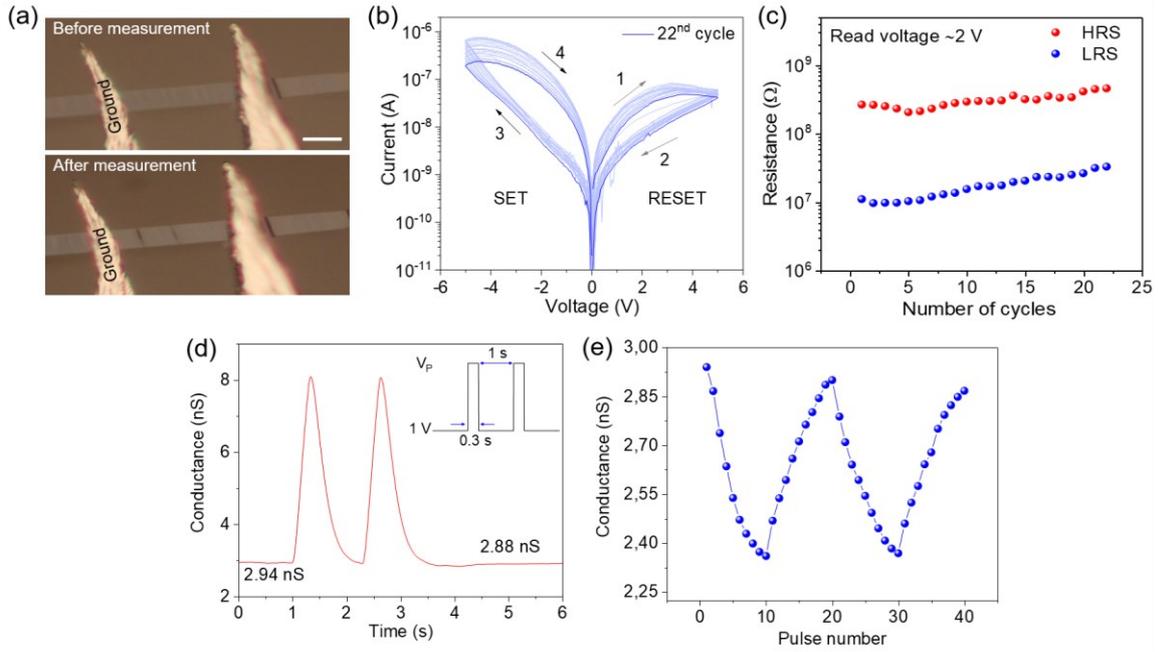

**Figure 5:** Resistive switching and synaptic behavior of $K_xWO_3$ devices. **(a)** Optical images of a crystal contacted with indium probes before and after measurement. Scale bar: 5 μm **(b)** Consecutive I–V cycles showing repeatable bipolar switching. **(c)** High- and low-resistance states (HRS/LRS) extracted at a read voltage of ~2 V. **(d)** Short-term depression (STD) under paired-pulse stimulation ($V_p$ = 4 V, 0.3 s, 1 s interval). **(e)** Long-term depression (LTD) and potentiation (LTP) triggered by trains of positive ($V_p$ = +5 V) and negative ($V_p$ = –5 V) pulses at a read voltage of 1 V.

**Conclusion**

To conclude, we developed an SLS growth approach to synthesize single-crystalline $K_xWO_3$ nanobelts with thicknesses down to ~36 nm and lateral sizes exceeding 100 μm. The crystals exhibit optically contrasting domains arising from inhomogeneous $K^+$ distribution formed during the synthesis. The application of external bias alters the K+ distribution, leading to the disappearance of these domains. Furthermore, IV sweeps reveal reproducible bipolar switching with resistance ratios of up to 30. Under pulsed excitation, the devices demonstrate both short-term and long-term synaptic plasticity, with switching energies of ~25 nJ. Overall, these results position $K_xWO_3$ as a model system for probing electric-field-driven alkali-ion migration in tunnel-structured oxides and highlight its potential as a stable and energy-efficient platform for neuromorphic device applications.

**Experimental Section**

**Sample Preparation:** $K_xWO_3$ crystals were synthesized by a confined SLS growth method. $WO_3$ powder (99.99%, Sigma-Aldrich) and KI (99%, Merck) were mixed in a 1:1 weight ratio, ground in an agate mortar, and dispersed on a c-cut sapphire substrate, which was capped with another sapphire to enforce confinement. The assembly was placed in a split-furnace CVD chamber, evacuated to $10^{-2}$ mbar, purged with high-purity Ar, and heated to 700 °C under 100 sccm of Ar flow.

**Characterizations and Transfer:** The morphology, composition, and crystal structure of $K_xWO_3$ crystals were characterized by OM (Olympus BX53), XRD (PANalytical equipped with Cu Kα radiation), AFM (Asylum Research MFP-3D), Raman spectroscopy (WITec Alpha300, 532 nm excitation), SEM (FEI Helios NanoLab DualBeam), TEM (FEI Tecnai G2 F30) equipped with SAED and EDX, and XPS (Thermo Scientific K-Alpha). For TEM sample preparation, poly(methyl methacrylate) (PMMA) was spin-coated on the crystal surface, and the PMMA-supported films were released by floating in deionized water and transferred onto Cu grids. After drying on a hot plate at 60 °C for 10 min, the PMMA layer was removed in hot acetone at 40 °C for 10 min.

**Device Fabrication and Measurements:** Two-terminal devices were fabricated by contacting the $K_xWO_3$ crystal with molten indium probes, a method established in our previous work[42]. Electrical transport and pulsed measurements were carried out using a Keithley 2400 source-meter and a Keysight B2901A unit.

## Author Contributions

A.A.S. and H.R.R. carried out the crystal synthesis and characterization with assistance from A.P. and D.P. Device fabrication and measurement were performed by A.A.S. and H.M.S., with additional support from A.P. and D.P. T.S.K. conceived the project and supervised the experiments. The manuscript was written by A.A.S., A.P., and T.S.K. with input from all authors.

## Data Availability

The data that support the findings of this study are available from the corresponding authors upon reasonable request.

## Conflict of Interest

The authors declare no competing financial interest.

## Acknowledgements

This work was supported by the Scientific and Technological Research Council of Türkiye (TÜBİTAK) under grant numbers: 120N885, 121F366, and 125F047.

# Synthesis of Ultra-thin Potassium Tungsten Bronze Single Crystals with Optically Contrasting Domains and Resistive Switching

Abdulsalam Aji Suleiman[1,2,3,4*], Amir Parsi[2], Hafiz Muhammad Shakir[5+], Hamid Reza Rasouli[2++], Doruk Pehlivanoğlu[2], Talip Serkan Kasırga[1,6*]

[1]National Nanotechnology Research Center – UNAM, Ankara 06800, Türkiye

[2]Institute of Materials Science and Nanotechnology, Bilkent University, Ankara 06800, Türkiye

[3]Department of Engineering Fundamental Sciences, Sivas University of Science and Technology, Sivas 58000, Türkiye

[4]Sivas Cumhuriyet University Nanophotonics Application and Research Center CÜNAM, Sivas 58140, Türkiye

[5]Department of Physics, Bilkent University UNAM, Ankara 06800, Türkiye

[6]Department of Physics, Middle East Technical University, Ankara 06800, Türkiye

[+]Current Address: College of Science and Engineering, Hamad Bin Khalifa University, Education City, Qatar Foundation 34110, Doha, Qatar

[++]Current Address: Institute of Physical Chemistry, Friedrich Schiller University Jena, 07743, Jena, Germany

*Corresponding author e-mail: suleiman@sivas.edu.tr, kasirga@unam.bilkent.edu.tr

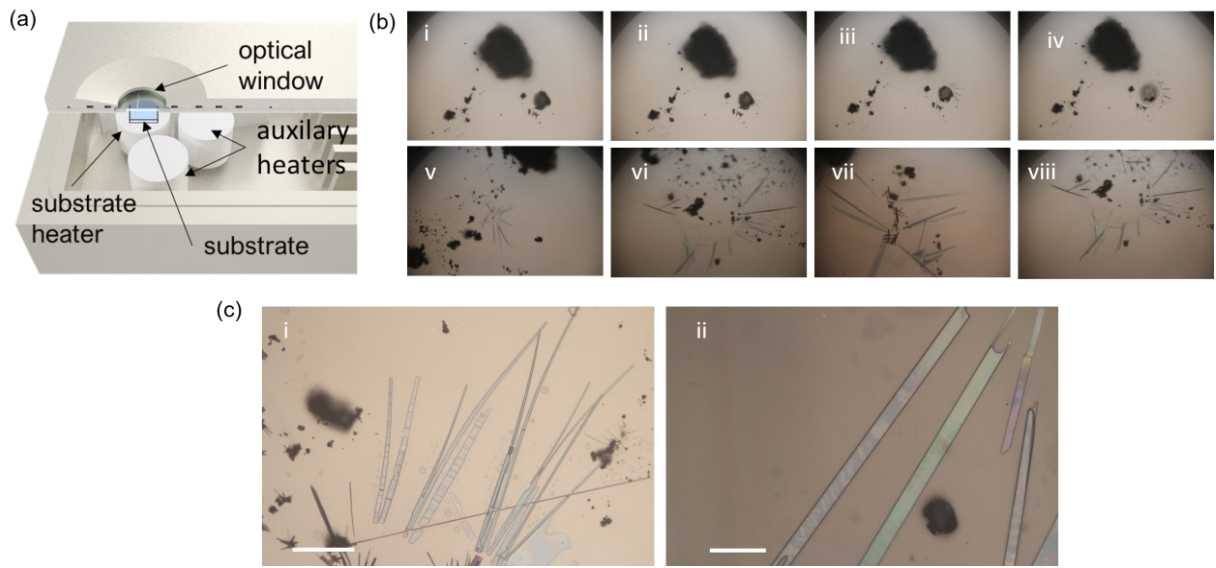

**Figure S1: (a)** Schematic of the real-time optical (RTO) CVD chamber. **(b)** Time-lapse optical images showing crystallization dynamics. **(c)** OM images of as-grown $K_xWO_3$ crystals with scale bars of 40 μm (i) and 20 μm (ii).

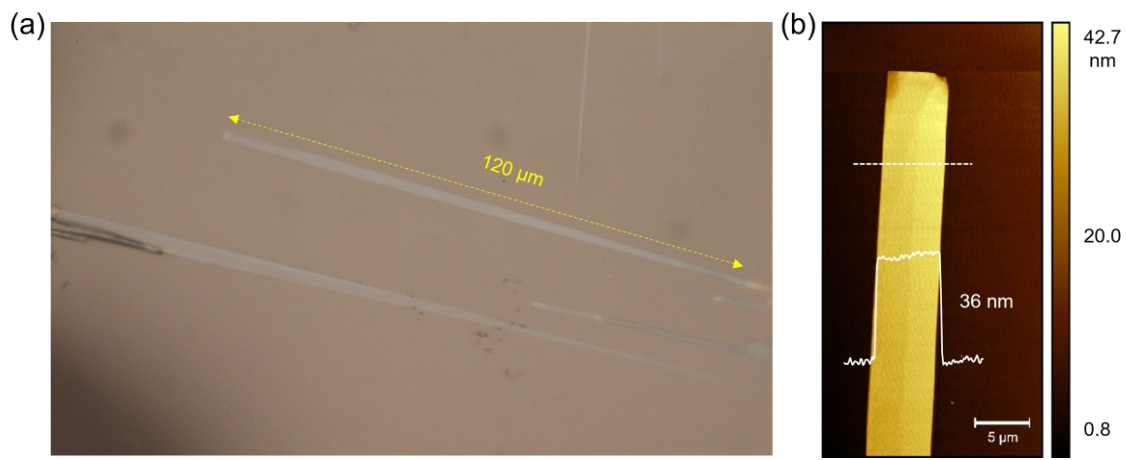

**Figure S2: (a)** OM image and **(b)** AFM topography of $K_xWO_3$ nanobelt.

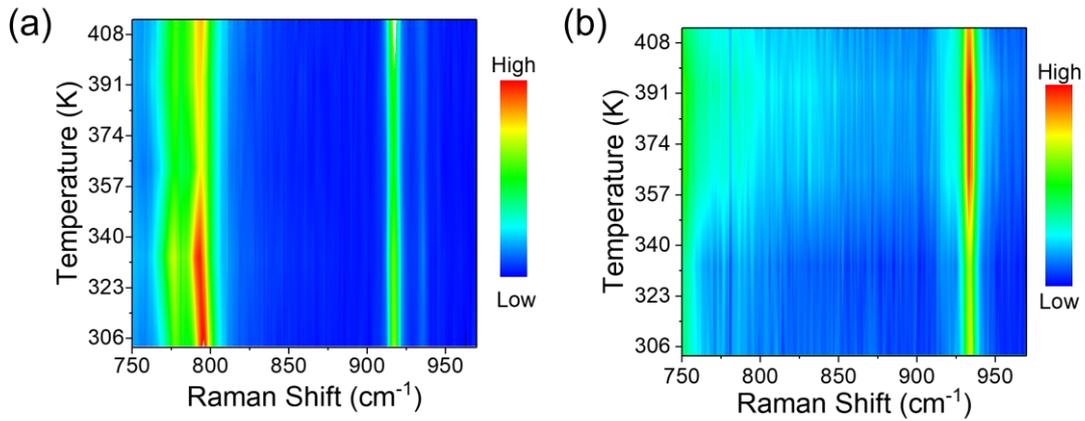

**Figure S3:** Temperature-dependent Raman spectra: contour maps of **(a)** bright and **(b)** dark domains in $K_xWO_3$ crystals.

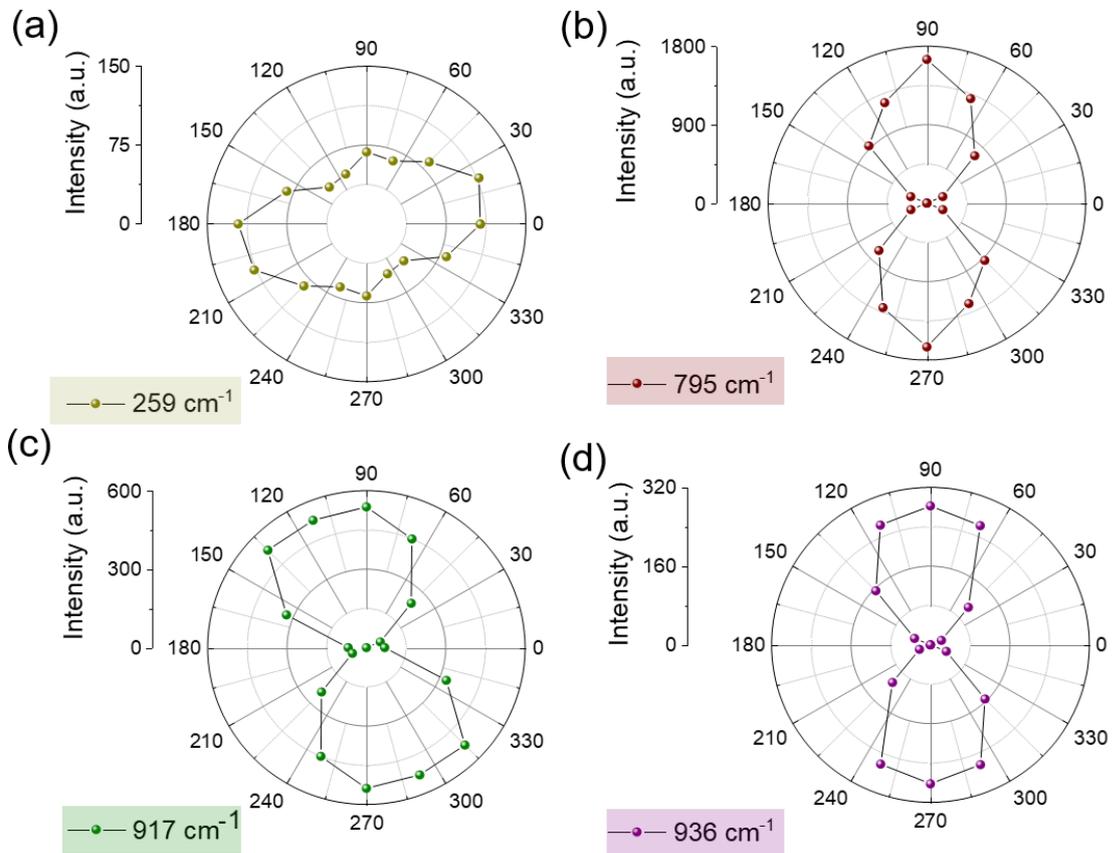

**Figure S4:** Angle-resolved Raman spectra of a $K_xWO_3$ nanobelt for the modes at **(a)** 259 cm$^{-1}$, **(b)** 795 cm$^{-1}$, **(c)** 917 cm$^{-1}$, and **(d)** 936 cm$^{-1}$.

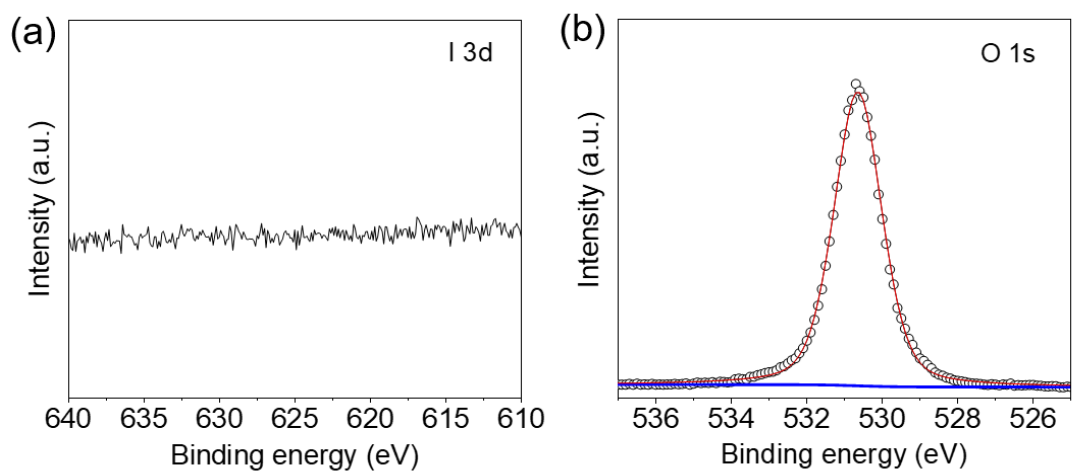

**Figure S5:** XPS spectra of **(a)** I 3d showing no detectable iodine signal, and **(b)** O 1s highlighting the contribution dominated by the sapphire substrate.

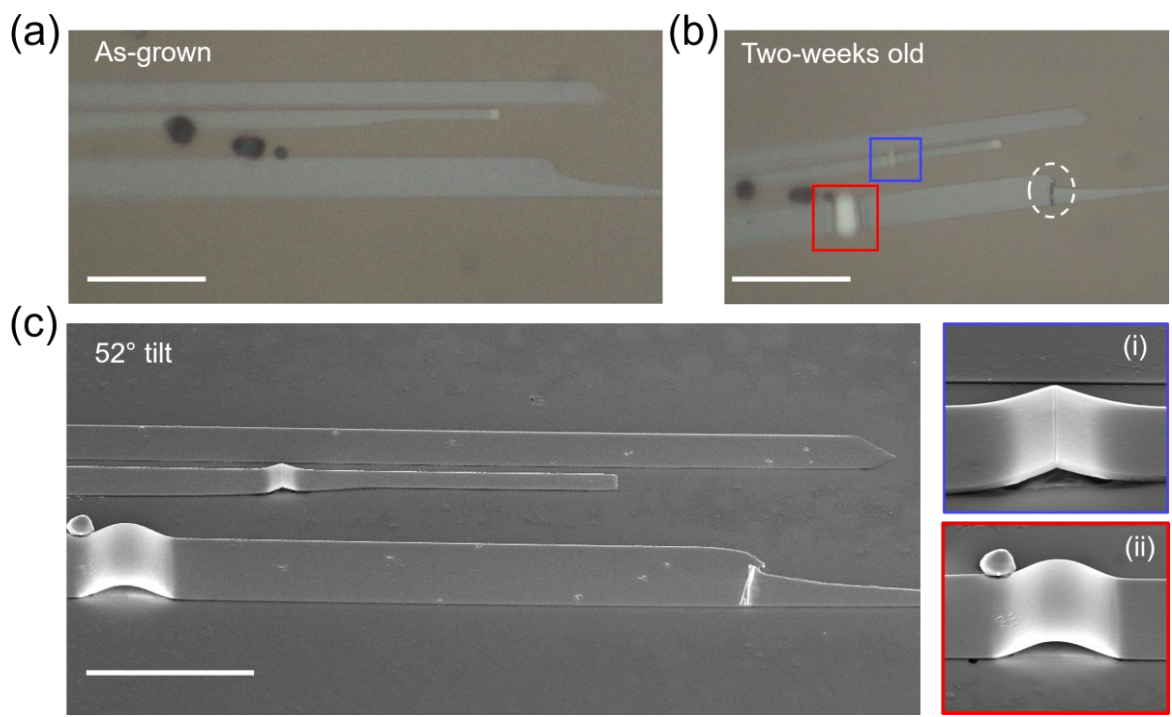

**Figure S6:** Aging-induced buckling in $K_xWO_3$ nanobelts. **(a)** OM image of an as-grown crystal. **(b)** OM of the same belt after two weeks at ambient conditions, showing the onset of out-of-plane buckling oriented along the c-axis. **(c)** SEM acquired at 52° tilt; (i) magnified view of the blue box in (b), and (ii) magnified view of the red box in (b).

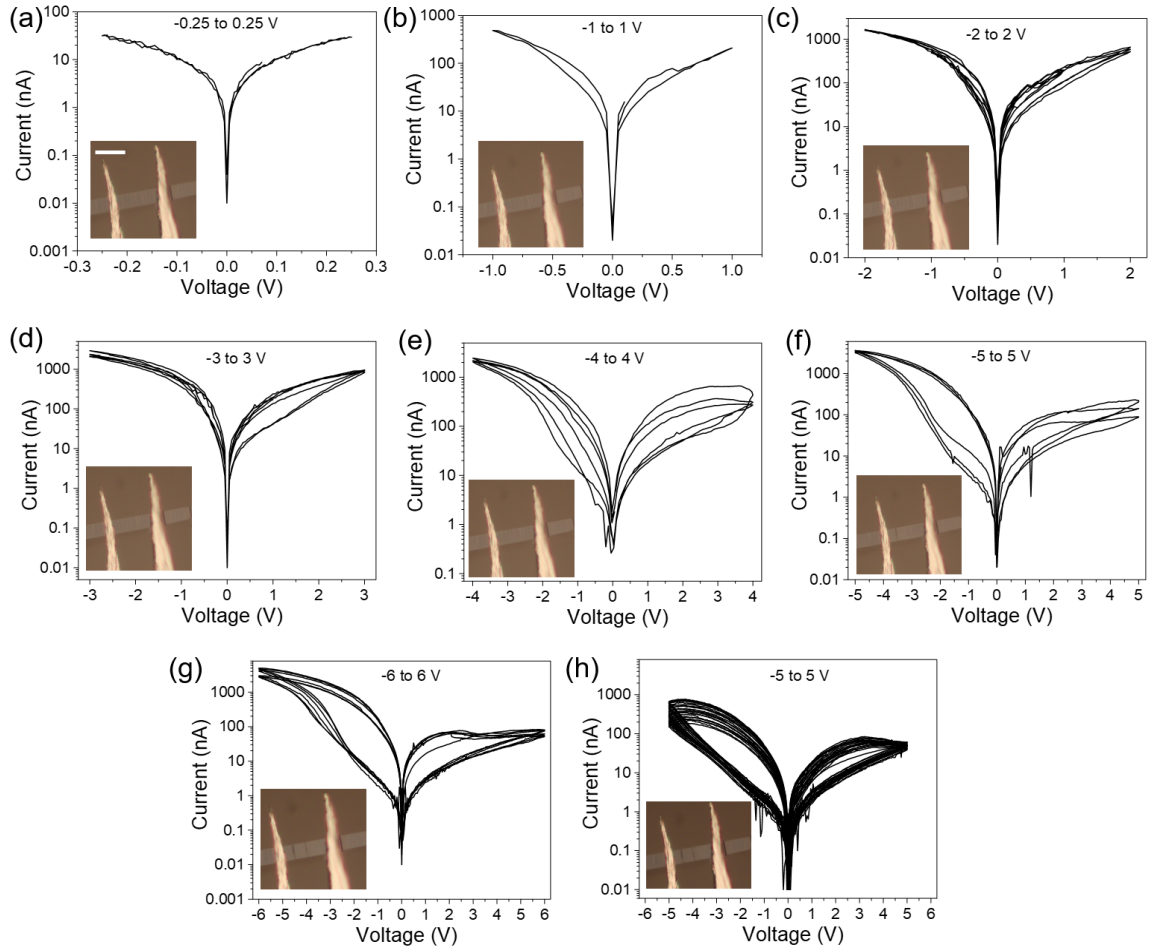

**Figure S6: (a-h)** Real-time evolution of consecutive I–V sweeps, illustrating the gradual stabilization of bipolar switching behavior in the $K_xWO_3$ device.

**Table 1:** Comparison of representative ion-electronic memristive devices.

| Materials | Ion pathway | Operating metrics (V, time, energy) | On/Off ratio & endurance | Mechanism | Refs. |
|---|---|---|---|---|---|
| Na–MnO$_2$ | Layered oxide; Na$^+$ intercalation / Mn redox | ~4–6 V; ~100 ms; ~0.5 nJ | Up to ~100; 60 pulse cycles | Field-driven structural phase transition layered ↔ spinel | 1 |
| K–MnO$_2$ | Layered birnessite; field-driven layered ↔ spinel transitions | ±8 V; 25 ms pulses (STP); 150 ms interpulse (LTP/LTD) | ≥5600 pulse endurance; ratio NR | Field-driven structural phase transition layered ↔ spinel | 2 |

| K–MnO$_2$ | Layered birnessite; interlayer K$^+$ migration | ~4–6 V; ~100 ms; ~0.5 nJ | ~10; limited cycles | Field-driven structural phase transition layered ↔ spinel | 3 |
| Cu$_2$S | Intrinsic Cu$^+$ migration | 0.1 V SET; ms-scale; ~1 μW @100 mV (power) | ~400; 500 pulse cycles | Ion-driven monoclinic ↔ tetragonal phase change | 4 |
| K$_x$WO$_3$ | K$^+$ redistribution in 1D channels | 3–5 V; 100–300 ms; 0.5–5 nJ | 10–30 (→ ~10 after cycling); ≥20 cycles | Mixed ionic–electronic conduction | This work |